\documentclass[12pt]{article}
\usepackage{latexsym}
\usepackage[dvips]{graphicx}
\usepackage[english]{babel}
\usepackage{epsfig}
\newcommand \Pomeron {I\!\!P}

\begin{document}
\begin{center}
{\Large {\bf
Onset of Perturbative Color Opacity at Small $x$ and $\Upsilon $
Coherent Photoproduction off heavy nuclei at LHC}} 
\vskip 0.5cm 
L.~ FRANKFURT\\
{\it Nuclear Physics Dept., School of Physics and Astronomy, Tel Aviv
  University, 69978 Tel Aviv, Israel}\\
\vskip 0.25cm 
V.~ GUZEY\\
{\it Institut f{\"u}r Theoretische Physik II, Ruhr-Universit{\"a}t Bochum,
  D-44780 Bochum, Germany}\\
\vskip 0.25cm
M.~ STRIKMAN\\
{\it Department of Physics, the Pennsylvania State University, State
  College, PA 16802, USA}\\
\vskip 0.25cm
M.~ZHALOV\\
{\it Petersburg Nuclear Physics Institute, Gatchina 188350, Russia}
\date{}
\end{center}

\begin{abstract}
We study photon-induced coherent production of $\Upsilon$ in 
ultraperipheral heavy ion collisions  at LHC and demonstrate 
that the counting rates will be sufficient to measure nuclear 
shadowing of generalized gluon distributions. This will establish
the transition from the regime of color transparency to the regime 
of perturbative color opacity in an unambiguous way. 
We argue that such measurements will provide the possibility to
investigate the interaction of  ultra-small color dipoles with nuclei
in QCD at large energies, which are beyond the reach of the electron-nucleon 
(nucleus) colliders, and will unambiguously discriminate between the leading
twist and higher twist scenarios of gluon nuclear shadowing.
\end{abstract}

\section{Introduction}

High energy coherent vector meson photo- and electroproduction
provides new tools  for understanding of many exciting phenomena
in the QCD physics of the hadron-nucleus interactions. The photon
wave function comprises a variety of hadronic components as well
as a direct contribution of the quark-antiquark pairs.
In the high energy regime, the photon transforms into a $q\bar q$ pair
long before the nuclear target.   
The transverse size of the produced $q\bar q$ pair is controlled by the mass 
of the quark ($r \propto  m_{q}^{-1}$, provided $m_q \gg \Lambda_{QCD}$).
Correspondingly, the scale of virtuality in the considered processes
is $Q^2 \ge  m_q^2$.
The strength of the high-energy interaction of small dipoles 
in QCD depends on the size of the area occupied by the
color field within the interacting objects:
the smaller the size, the weaker the interaction~\cite{Low,Gunion}. 
Probably the most sensitive indicator  of the size of the interacting objects
is the $t$-dependence of exclusive vector meson electro- and photoproduction.
 The current HERA data on the vector meson electroproduction on
the proton target, see e.g.~ \cite{H1}, are consistent with the prediction 
of~\cite{BFGMS} that the t-slopes of 
the $\rho$ and $J/\psi$ production production cross sections 
should converge to the same value with the rate of convergence consistent with 
the estimate of~\cite{FKS}.
This indicates that at small $x$, $x=(M_{V}^2+Q^2)/s$ ($x$ is related to the 
momentum fractions of the exchanged gluons -- see the discussion of Eq.~(\ref{eq4});
 $s$ is the invariant energy for $\gamma - N$ scattering), the
configurations of much smaller size than the average light meson size
dominate  $J/\psi$ production for all $Q^2$ and  $\rho$ production
for $Q^2 \ge 5$ GeV$^2$.  
In the kinematics of  small $x$ and large $Q^2$, the QCD factorization theorem 
predicts that the cross sections of coherent production of vector mesons off
nuclei are proportional to the square of the nuclear gluon parton density
\footnote{Note that the corrections due to the skewedness effects
at large $Q^2$ and small $x$  are calculable in terms of the 
QCD evolution equation for the generalized parton distributions.}
 $G_A(x,Q^2)$~\cite{BFGMS,Collins,FMS}.
  It is reasonably well known from various data analyses that  $G_A(x,Q^2)$
is not shadowed and even may be enhanced at  $0.02\div 0.03\leq x\leq 0.2$:
$G_{A}(x,Q^2)\geq AG_{N}(x,Q^2)$. 
One should note that the small-$x$ behavior of nuclear parton distributions
 is crucial for various aspects of small-$x$ dynamics with nuclei.
 The nuclear parton distributions have been analyzed both phenomenologically,
 by fitting to the available data~\cite{Eskola,Kumano}
(which for $Q^2\ge 2$ GeV$^2$ are available only for 
$x \ge 10^{-2}$), and  theoretically, by using various models, see 
Ref.~\cite{guide} for the review. 
These analyses suggest a rather significant nuclear shadowing assuming that 
the leading twist effects dominate. However, the  $F_{2A}/F_{2D}$
data are restricted to rather moderate $Q^2$, where higher twist effects
due to e.g. diffractive vector meson production are not negligible.

On the theoretical end, there are two types of models. 
One class focuses on sufficiently  large virtualities,
where leading twist dominates.
It assumes that the soft physics (aligned jet model)
which dominates at the starting scale of evolution for quark shadowing should be present in the gluon sector as well, so 
that  the gluon shadowing effects at the initial evolution 
scale of a few GeV$^2$ should be comparable 
to that of quarks~\cite{FLS,Eskola}.
More recently a more quantitative approach became possible. It is
based on the combination of the Gribov theorem connecting diffraction and
shadowing~\cite{Gribovinel}, and 
the Collins factorization theorem for hard diffraction in DIS~\cite{FS99}.
As a result, this approach  effectively takes into account the nonconservation of the
number of particles in  QCD evolution and the restoration of the
gluon fields by the small color singlets. When combined with  the HERA  hard 
diffractive  data and quasieikonal modeling of the multiple rescattering terms,
it leads to the 
prediction that the gluon shadowing is significantly larger than
that for quarks at the starting scale of QCD evolution of $Q^2 \sim 4$ GeV$^2$.

Another group of the models~\cite{RL} aims to model the effects of the
modification of 
nuclear gluon field both in the  regime of high gluon densities, where 
the decomposition over the twists is not applicable and at large $Q^2$ where the leading twist dominates. It assumes that parton 
densities are saturated at $Q^2\le Q^2_s$ ($Q_s^2(x)$ is the saturation scale).
An important feature of these models is that they do not lead to the leading 
twist shadowing at  $Q^2\gg Q^2_s$, see e.g. discussion in~\cite{Mueller,Kovchegov}.
In particular, in these models  the effect of nuclear shadowing for the interaction 
of a small dipole is modeled in the frozen impact parameter eikonal approximation 
with the dipole coupled to the unscreened nucleon gluon field, leading to the
shadowing $\propto  xG_N(x,Q^2)/Q^2$.

A large  leading twist gluon shadowing would significantly slow down
the increase of the cross section of the small dipole-nucleus interaction
with energy (which is proportional to the gluon density) and, hence,
 would significantly extend the $x$ range, where 
unitarity for the interaction of small dipoles with nuclei is
not violated in the DGLAP approximation. As a result, this may  
strongly affect the pattern of the possible onset of the regime of the 
black body limit (BBL).

In the $x$-range, where nuclear gluons are not shadowed,
 one expects the regime of color transparency for coherent
production of onium states. A leading twist shadowing would lead
to the onset of the perturbative color opacity with the
 decrease of $x$. However, this shadowing merely slows down
 the increase of the 
gluon density with the decrease of $x$. Hence, at sufficiently small $x$,  
 the perturbative color opacity regime will be also violated.
This may lead to the onset of BBL or another nonlinear regime of QCD.

The crucial questions for understanding 
 the interplay among the above mentioned
 phenomena are the following.
What is the kinematical region where the squeezing of the $q\bar q$ dipole
takes place, which justifies the application of perturbative QCD?
What are the leading mechanisms of nuclear shadowing at very small $x$, where 
the small size dipole-nucleon interaction 
become large?
Can the gluon saturation effect at small $\it x$  stop the increase
of the cross section before reaching the BBL. 

In order to  establish the effect of nuclear opacity and
 clarify the role of the leading and higher twist mechanisms of  nuclear shadowing 
in an unambiguous way, it is necessary to investigate the
interaction of a small dipole with the nuclear  medium and check
whether the amplitude of the interaction is reduced as compared to the
color transparency expectation of the amplitude being proportional to
the atomic number.  Production of light vector mesons at small $x$ in DIS 
by the longitudinally polarized photons and photoproduction of onium 
states seem to be optimal for these purposes. The studies of the dipole 
model in Ref.~\cite{martin} suggest that one is safely into the perturbative
domain at $x\le 10^{-3}$ only if the transverse distances are below 
$d \sim 0.3$ fm. For the light mesons this requires $Q^2 \ge 15$ GeV$^2$
and, hence, an eA collider with energies of HERA and large luminosity. 
The photoproduction of $J/\psi$ appears to be the borderline case -- the
amplitude seems to be getting a significant contribution from the 
transitional region between perturbative and nonperturbative QCD.
Although  $J/\psi$ photoproduction suits well the studies of the 
onset  of color opacity, in general it may be rather problematic to distinguish the 
perturbative regime, where the amplitude is proportional to the gluon density, 
from the regime, where the fast increase of the gluon density results in the 
violation of the leading twist approximation and the transition to the black 
body limit (BBL) takes place. In Ref.~\cite{FSZrho} it was
 suggested to study the onset of BBL in the coherent 
photoproduction of dijets from heavy nuclei since in this case 
the role of the leading twist shadowing is negligible.
On the contrary,  the coherent $\Upsilon$ photoproduction off heavy nuclei 
seems to be essentially unique for the investigation of the onset of the perturbative
color opacity in the leading twist (LT) shadowing regime. In this case, 
the interaction  is dominated by such small transverse interquark 
distances $d_t$, $d_t \sim 0.1$ fm,
 that physics of interactions is definitely perturbative up 
to very large energies. For example, the cross section of the 
interaction of a dipole with $d_{t}\sim 0.1$ fm is
$\sigma(d_t=0.1\,{\rm fm}, x=10^{-4})=3.5$ mb
 and is even smaller for larger $x$.

 In order to
probe the onset of perturbative color opacity in the $\Upsilon$ production,
one would need to reach $x \ll 10^{-2}$. This would require an
 electron-nucleus collider in the HERA energy range with a luminosity
per nucleon comparable to the luminosity in the electron-proton collisions,
 which appears to be rather problematic for HERA.
At the same time there are plans  for studying the photoproduction
off nuclei  at LHC using ultraperipheral heavy ion collisions,
including the studies of the  coherent onium photoproduction.
For the review and extended list of references, see Ref.~\cite{upc}.

Recently it was demonstrated~\cite{FSZpsi} that  the yield 
of $J/\psi$ in the coherent production 
in the kinematics of the ultraperipheral ion collisions at LHC  
is suppressed at the central rapidities by a factor of 6
compared to that in the impulse approximation. Such a  significant 
effect of the suppression would clearly signal  the revealing of 
the color opacity phenomena. However, as mentioned above, 
the gluon virtualities in this case are on the borderline between 
the perturbative and nonperturbative QCD regimes 
($Q^2_{eff}\sim 3\div 4$ GeV$^2$) and  the leading twist approximation 
might not be accurate enough for the evaluation of the
scattering amplitude, especially since the
nonperturbative region  could probably give a significant 
correction~\cite{martinpsi}.

In summary,
 $\Upsilon$ photoproduction stands out as the simplest probe 
of the propagation of small color dipoles, $d_t=0.1$ fm, at LHC
energies, which will allow to investigate the nuclear gluon fields at the 
effective scale  $Q^2 \sim 40$ GeV$^2$. In this case,  the estimates
of  nuclear shadowing via the leading twist mechanism based on 
perturbative QCD used in Ref.~\cite{FSZpsi}, will be much better justified. 
The main objective of this work is to
present the results of  a study of the coherent production
of $\Upsilon$ in the ultraperipheral collisions of heavy ions in 
the kinematics of LHC.

\section{Production of $\Upsilon $ in ultraperipheral collisions}

   The  cross section of the coherent $\Upsilon$ production
integrated over the transverse momenta of the nucleus, which emitted a
photon,  can be written using the standard Weizsacker-Williams 
approximation~\cite{ww} in the  form:
\begin{equation}
{d \sigma(AA\to \Upsilon AA)\over dy}=
{N_{\gamma}(y)}  \sigma_{\gamma A\rightarrow \Upsilon A}(y)+
{N_{\gamma}(-y)}  \sigma_{\gamma A\rightarrow \Upsilon A}(-y) \,.
\label{base}
\end{equation}
Here 
$y=ln{\frac {2\gamma m_{N}M_V} {s}}$ is  the c.m. rapidity of the produced $\Upsilon$;
$\gamma$ is Lorentz factor;
$\sigma(y)$ is the cross section for the photoproduction of $\Upsilon$;
${N_{\gamma}(y)}$ is the
 flux of the equivalent photons~\cite{flux}:
\begin{equation}
N(y)=\frac {Z^2\alpha} {{\pi}^2}\int d^2b \Gamma_{AA}({\vec b})
 \frac{1} {b^2}X^2
\bigl [K^2_1(X)+\frac {1} {\gamma} K^2_0(X)\bigr ]\,,
\end{equation}
where $K_0(X)$ and $K_1(X)$ are modified Bessel functions of the argument 
$X=bm_V e^y/(2\gamma)$;  
${\vec b}$ is the transverse distance between the centers of the
colliding nuclei. The standard Glauber profile factor, 
\begin{equation}
\Gamma_{AA}({\vec b})=\exp\biggl (-\sigma_{NN}
\int \limits^{\infty}_{-\infty}dz\int d^2b_1
\rho_A(z,{\vec b_1})\rho_A(z,{\vec b}-{\vec b_1})\biggr ) \,,
\end{equation} 
accounts for the inelastic strong interactions of
the nuclei at impact parameters  $b \le 2R_A$ and, hence, suppresses
the corresponding contribution to the $\Upsilon$ photoproduction.
In our calculations we use the nuclear matter density
$\rho_A(z,{\vec b})$ obtained from the mean field
Hartree-Fock-Skyrme (HFS) model, which describes  many global properties
of nuclei~\cite{hfs} as well as  many single-particle nuclear structure
characteristics   extracted from
the high energy $A(e,e \prime p)$  reactions~\cite{lsfsz}.

\begin{figure}
\begin{center}
\includegraphics[scale=.8]{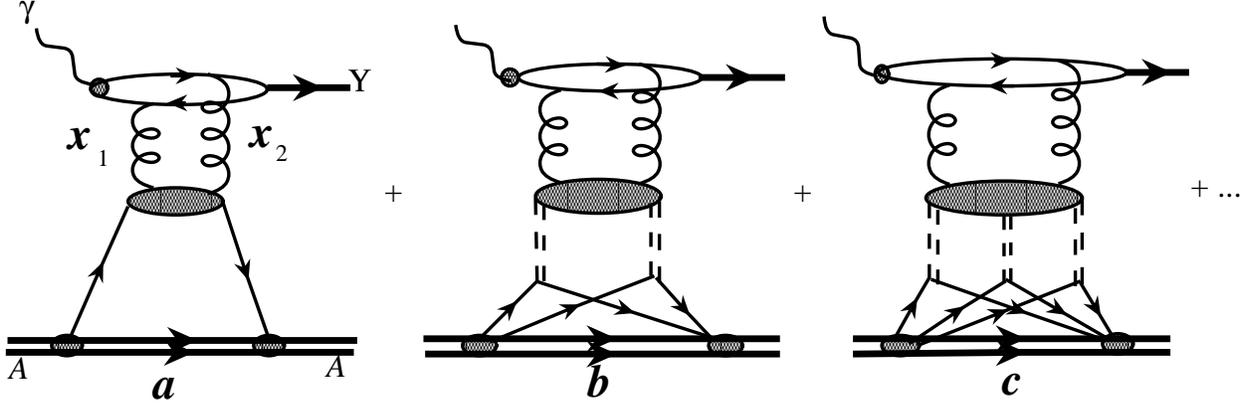}
\caption{High energy quarkonium photoproduction in the leading
twist approximation.}
\end{center}
\label{ltdiag}
\end{figure}

The amplitude of the $\Upsilon $ photoproduction
(necessary for the calculation of  $\sigma_{\gamma A\rightarrow
  \Upsilon A}$ in Eq.~(\ref{base})) in the leading twist approximation 
is described by the series of the Feynman diagrams in Fig.~\ref{ltdiag}.
The QCD factorization theorem for  exclusive meson 
photoproduction~\cite{Collins,FMS,BFGMS} allows one to express the 
imaginary part of the forward amplitude for the production of a heavy vector 
meson by a  photon, $\gamma + T \to V + T$,
through the convolution of the wave function of the meson at the zero 
transverse separation between the quark and antiquark, the hard 
interaction block and the generalized parton distribution (GPD) of the target,
 $G_{T}(x_1,x_2,Q^2,t_{min})$, at $t_{min}\approx -x^2m_N^2$.  
The gluon light cone fractions $\it x_i$
for the gluons attached to the quark loop satisfy the relation:
\begin{equation}
x_1-x_2={m^2_{\Upsilon}\over s}\equiv x \,,
\label{eq4}
\end{equation}
where $s=4E_N \omega=4\gamma \omega m_N$
is the invariant energy for $\gamma - N$
scattering ($E_N=\gamma m_N$ is the energy per nucleon
in the c.m.  of the nucleus-nucleus  collisions).
If the  quark Fermi motion and binding effects were negligible,
$x_2\ll x_1$. Numerical estimates using realistic potential model wave
functions indicate that for  $J/\psi$, $x_1\sim 1.5 x, x_2\sim x/2$
\cite{martinpsi}, and that for $\Upsilon$, $x_2/x_1\sim 0.1$ \cite{fmsupsilon}.
Modeling of the GPDs at moderate $Q^2$ suggests that, to a good
approximation, 
$G_T(x_1,x_2)$
 can be approximated by the gluon density
at $x=(x_1+x_2)/2$ \cite{BFGMS,Rad}. For large $Q^2$ and small $x$,
GPDs  are dominated by the evolution from  $x_i(init) \gg x_i$. 
Since the evolution conserves $x_1-x_2$, the effect of skewedness
 is determined primarily by the evolution from nearly diagonal  distributions.

Hence, we can approximate  the
amplitude of the $\Upsilon$ photoproduction off nucleus at $k_t^2=0$, 
presented by a series of diagrams in Fig.~\ref{ltdiag}, as
\begin{equation}
M(\gamma +A \to \Upsilon + A)=M(\gamma +N \to \Upsilon + A)
{\frac{G_A(x,Q^2_{eff})} {AG_N(x,Q^2_{eff})}} F_A(t_{min})\,,
\label{amplitude}
\end{equation}
where $F_A$ is nuclear form factor normalized so that $F_A(0)=A$;
$Q^2_{eff}(\Upsilon)\sim 40$ GeV$^2$ according to the 
estimates of~\cite{FKS}. It is worth noting that really we are not
 sensitive here to the precise value of $Q^2_{eff}$ since for large $Q^2$,
 the gluon shadowing decreases slowly with $Q^2$.
Note that the skewedness effects, which are expected to be small
at $Q^2\sim$ a few GeV$^2$, become important in the $\Upsilon$ case.
They appear to  increase the cross
section of the  reaction $\gamma +p \to \Upsilon +p$ by a factor $\sim
2$, see Refs.~\cite{fmsupsilon,Martin:1999rn}. Potentially,  this  
could obscure the connection between the perturbative color opacity effect 
and shadowing of the nuclear gluon densities. However, the
  analysis of~\cite{fgsrev} shows that the ratio of GPD on nucleus and on a
nucleon for $t=t_{min}$ is a weak function of $x_2$,
slowly dropping from its diagonal value ($x_2=x_1$) with the decrease
of $x_2$. Overall this observation is in a agreement with the general
trend mentioned above that it is more appropriate to do a comparison of the
diagonal and non-diagonal cases at $x=(x_1+x_2)/2$.

The effect of leading twist nuclear shadowing can be quantified by
 considering
the diagram (b) in Fig.~\ref{ltdiag}. All possible unitary cuts of the 
diagram lead to inelastic shadowing in hadron-nucleus total cross 
sections~\cite{Gribovinel}.  Applying these ideas to DIS on nuclei
 and using the factorization theorem for hard
 diffraction~\cite{Collins},
 coupled with the QCD analysis of the HERA data on diffractive 
DIS~\cite{H1:1994},
nuclear parton distributions can be predicted~\cite{FS99,guide}.
In general, the theory of leading twist nuclear shadowing should 
be applied in two steps.   
Firstly, the nuclear gluon density distribution 
 is calculated at the starting evolution scale of the 
effective momentum transfer $Q_{0}^2=4$ GeV$^2$ 
\begin{eqnarray}
&& {G_{A}({\it x},Q_{0}^2)}= {AG_{N}({\it x},Q_{0}^2)}-
8\pi \Re e \biggl [
{\frac {(1-i\eta )^2} {1+{\eta}^2}}
\int d\,^2\,b\, \int \limits_{-\infty}^{\infty} dz_1
\int \limits_{z_1}^{\infty}dz_2\int \limits_{\it x}^{{\it x}_{\Pomeron}^{0}}
d{\it x}_{\Pomeron} \nonumber \\
&&\times g^{D}_{N}(\frac {\it x} {{\it x}_{\Pomeron}},{\it x}_{\Pomeron},
Q_{0}^2,t_{min})\rho_{A}({\vec b},z_1)\rho_{A}({\vec b},z_2)
e^{i{\it x}_{\Pomeron}
m_N(z_1-z_2)}\,e^{-{\frac {1} {2}}\sigma_{eff}(x,Q_{0}^2)(1-i\eta )\int 
\limits_{z_1}^{z_2} dz\rho_A({\vec b},z)} \biggr ] \,.
\label{eik}
\end{eqnarray}
Here $g^{D}_{N}(\frac {\it x} {{\it x}_{\Pomeron}},
{\it x}_{\Pomeron},Q_{0}^2,t_{min})$
is the diffractive gluon density distribution of
the nucleon; 
${\it x_{\Pomeron}}$ is the Pomeron momentum fraction;
 $\eta$ is the
ratio of the real to imaginary parts of the elementary diffractive amplitude;
${\it x}_{\Pomeron}^{0}=0.03$ is the cut-off parameter of the theory.
As an input, we used the H1 parameterization~\cite{H1:1994} of
 $g^{D}_{N}(\frac {\it x} {{\it x}_{\Pomeron}},{\it x}_{\Pomeron},
Q_{0}^2,t_{min})$.
The effect of the interactions with three and more nucleons
of the target (graph (c) in Fig.~\ref{ltdiag}) is included in the 
attenuation factor $T=\exp(-\sigma_{eff}(x,Q_{0}^2)(1-i\eta )/2 
\int \limits_{z_1}^{z_2} dz\rho_A({\vec b},z))$.
The  effective cross section $\sigma_{eff}(x,Q_{0}^2)$ accounts for the
elastic rescattering of the produced diffractive state off the nuclear nucleon
in the series of the rescattering diagram 
(graphs (b) and (c) in Fig.~\ref{ltdiag}) and is defined by the equation
\begin{eqnarray}
\sigma_{eff}(x,Q_{0}^2)=\frac {16\pi } {(1+{\eta}^2)G_N(x,Q_{0}^2)}
\int \limits_{\it x}^{{\it x}_{\Pomeron}^{0}}
d{\it x}_{\Pomeron}
g^{D}_{N}(\frac {\it x} {{\it x}_{\Pomeron}},{\it x}_{\Pomeron},
Q_{0}^2,t_{min}) \,.
\end{eqnarray}
Secondly,
 since the photoproduction of $\Upsilon$ corresponds to larger $Q^2$ scales 
$\approx 40$ GeV$^2$, the gluon density distributions were evolved up to
this scale using  the NLO QCD evolution equations.  
In the limit of low nuclear density, when the effect of the attenuation
 factor $T$ can be neglected, Eq.~(\ref{eik}) can be applied to
 evaluate $G_A$ at any scale $Q^2$ and not only at $Q_0^2$. In the general 
case, Eq.~(\ref{eik}) should be used only at rather moderate $Q_0^2$ 
and should serve as an input for QCD evolution: The modelling of multiple
 rescattering by the attenuation factor $T$ ignores the fluctuations in
 the strength of the interaction around the average value $\sigma_{eff}$, 
which is a good approximation only at moderate $Q_0^2$.
Indeed, at $Q_0^2\approx 3\div 4 $ GeV$^2$, the interactions are 
predominantly soft with small cross section fluctuations.
The effect of cross section fluctuations is automatically included in 
 QCD evolution, which leads to violation of the
Glauber - like structure of the expression for the shadowing 
for $Q^2> Q_0^2$.

One can also address the question of the accuracy of 
the substitution of the ratio of  the generalized gluon densities by the
ratio of the diagonal parton densities  at the normalization scale. 
In the case of scattering off two nucleons (only graph (b) in Fig.~\ref{ltdiag}),
one  can express the leading twist screening via the nondiagonal matrix element of the
diffractive distribution function, $\tilde{g}^D$ (an analog of generalized PDF). 
It depends on the light-cone fraction, which the nucleon lost in the 
 $\left|in \right>$ and $\left<out\right|$ states:
  $x_{\Pomeron}$ $x_{\Pomeron}-x$ respectively,  $\beta_{in}=x_1/x_{\Pomeron},
\beta_{out}=(x_1-x)/(x_{\Pomeron}-x)$,  and $t,Q^2$.
If we make a natural assumption that 
\begin{equation}
\tilde{g}^D(x_1,x,x_{\Pomeron},Q_0^2,t)=
\sqrt{g^D(\beta_{in},Q^2_0,x_{\Pomeron},t)g^D(\beta_{out},Q^2_0,
x_{\Pomeron}-x,t)} \,,
 \end{equation}
we find that in the kinematics we discuss, the resulting
skewedness effects are small numerically.
They are smaller than the uncertainties in the input gluon diagonal  diffractive PDFs. 

\begin{figure}
\includegraphics[scale=0.8]{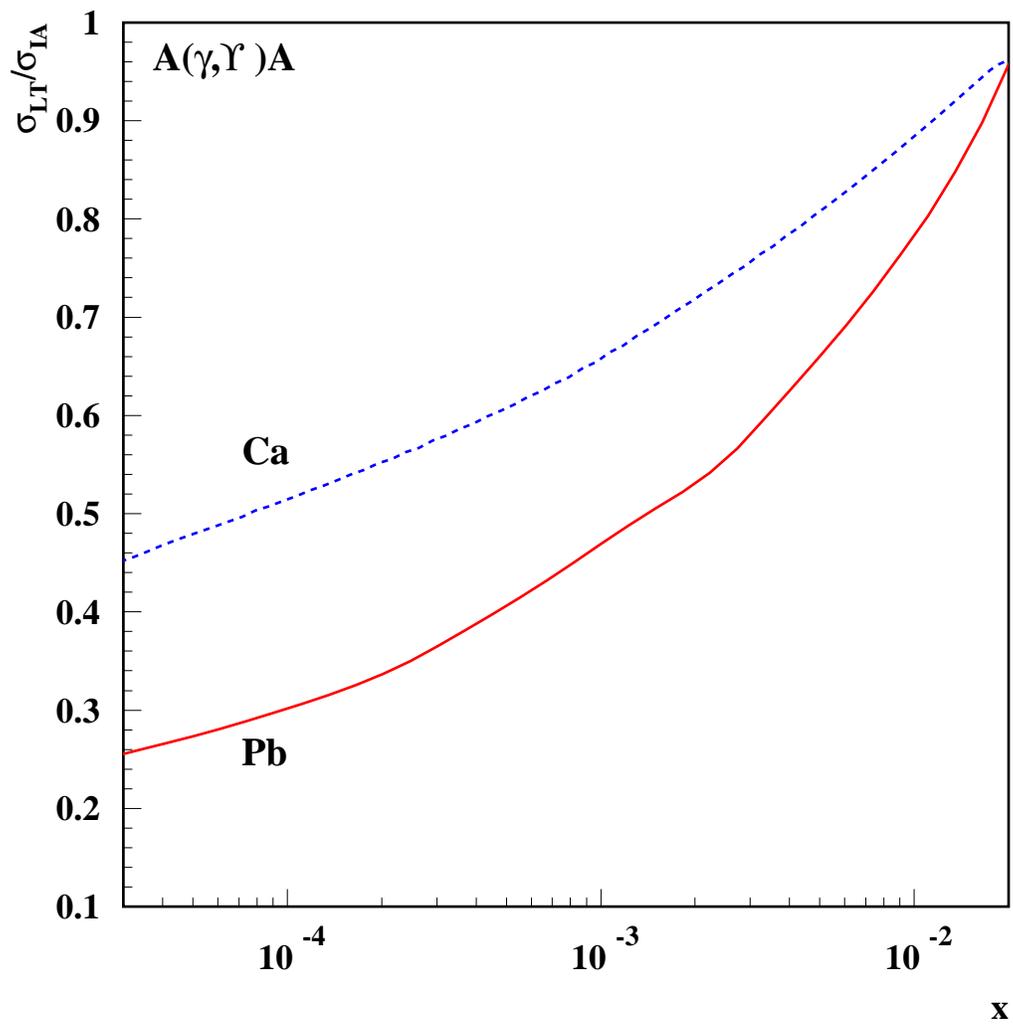}
\caption{The shadowing effect in the 
$\Upsilon$ photoproduction off Pb and Ca.}
\label{shad}
\end{figure}

The discussion of the nuclear gluon distribution culminates in the fact that  
the cross section of the process $\gamma A\rightarrow \Upsilon A$
can be readily obtained by squaring the amplitude in Eq.~(\ref{amplitude})  
\begin{equation}
\sigma_{\gamma A\rightarrow \Upsilon A}(s)=  
{d \sigma_{\gamma N \to \Upsilon N}(s,t_{min})\over dt}
\Biggl [\frac {G_{A}(\frac {M_{\Upsilon}^2} {s},Q_{eff}^2)}
{AG_{N}(\frac {M_{\Upsilon}^2} {s},Q_{eff}^2)}\Biggr ]^2
\int \limits_{-\infty}^{t_{min}} dt
{\left|
\int d^2bdz e^{i{\vec q_t}\cdot {\vec b}}
e^{-iq_{l}z}\rho_A ({\vec b},z)
\right|
}^2 \,,
\label{phocs}
\end{equation}
where $-t=|\vec{q}_t|^2+|q_l|^2$ is the square of the vector meson
transverse momentum and $q_{l}=m_{N}M_{\Upsilon}^2/s$ is the minimal 
longitudinal momentum transfer in the photoproduction vertex.

The elementary cross section of forward $\Upsilon$ photoproduction
in the energy range of interest is not known well experimentally.
In our calculation, we used a simple parametrized form
\begin{equation}
 {d \sigma_{\gamma N\to \Upsilon N}(s,t)\over dt}=10^{-4}
B_{\Upsilon}
\biggl [\frac {s} {s_0}\biggr]^{0.85} \exp(B_{\Upsilon}t)
\label{eq:cs}
\end{equation}
with the reference scale $s_0=6400$ GeV$^2$; the slope parameter
 $B_{\Upsilon}=3.5\,GeV^{-2}$ 
and the energy dependence, which follows from the 
calculations~\cite{fmsupsilon}
of the photoproduction of $\Upsilon$ in the leading $\log Q^2$ approximation 
with an account for the skewedness of the partonic density distributions.
The cross section is normalized so that the total
cross section is in $\mu b$. 

\begin{figure}
\includegraphics[scale=0.8]{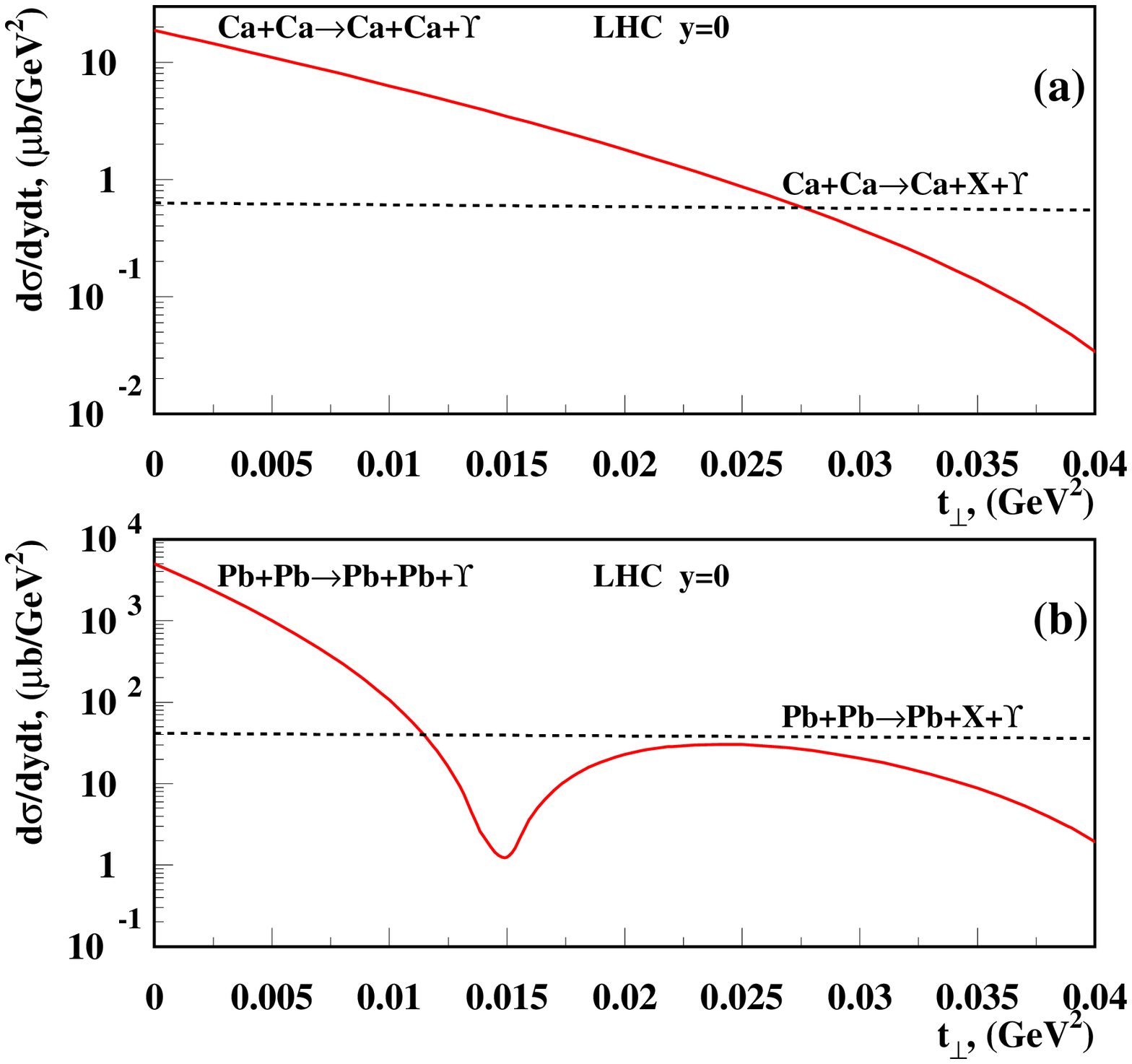}
\caption{The momentum transfer dependence
in the coherent $\Upsilon$ production in the 
UPC at LHC. 
The solid curve corresponds to the coherent cross section with accounting 
for nuclear shadowing; the dashed curve corresponds to incoherent $\Upsilon$ production.}
\label{dtyps}
\end{figure}

\begin{figure}
\includegraphics[scale=0.8]{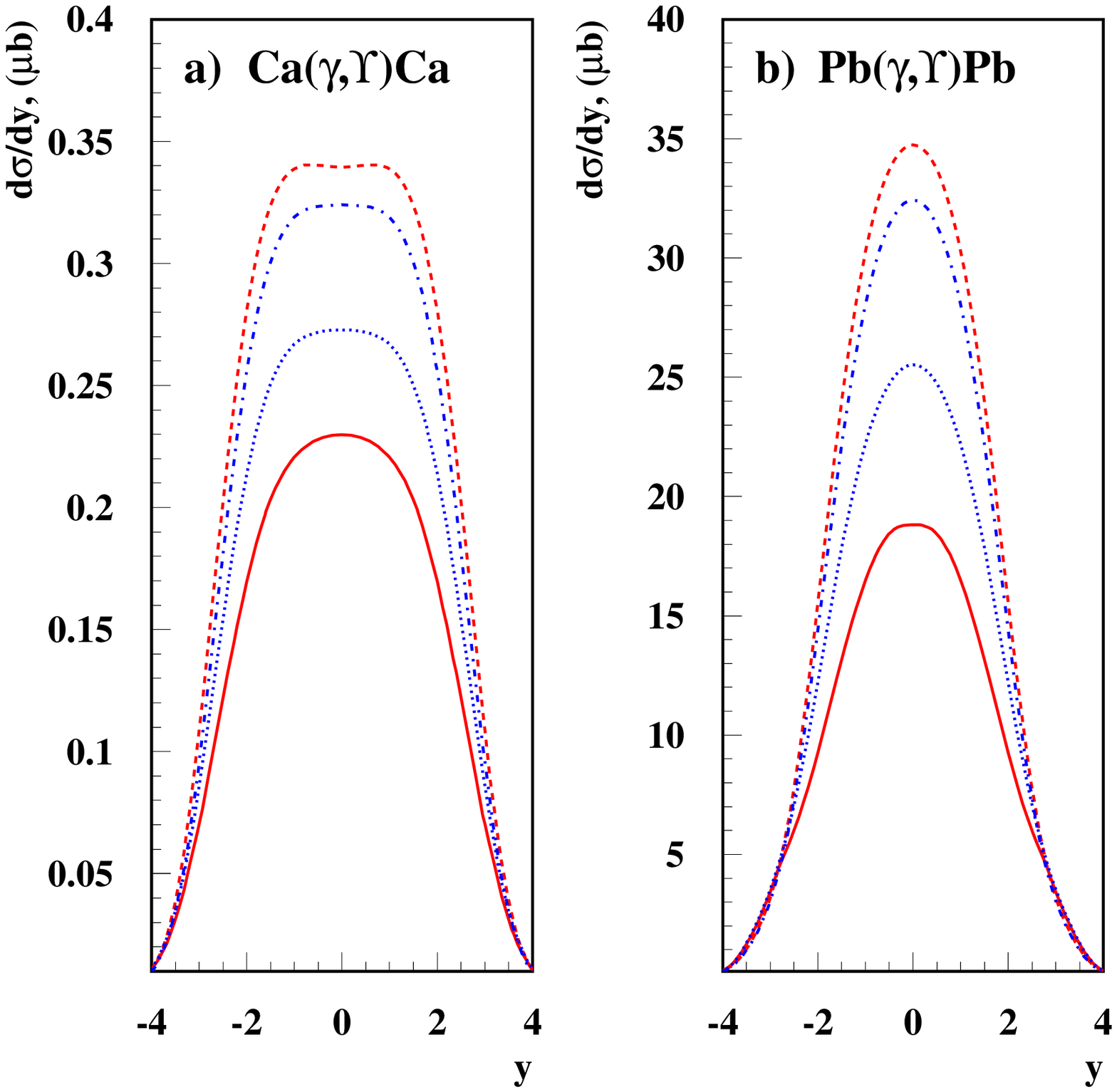}
\caption{The rapidity distribution for the coherent
$\Upsilon$ production in Ca-Ca and Pb-Pb in UPC at LHC. 
The solid curve corresponds to the calculation including leading twist nuclear
shadowing; the dotted curve corresponds to the calculation with the model of shadowing
of Eskola {\it et al.}; the dot-dashed curve is the calculation in the eikonal dipole rescattering 
model; the dashed curve corresponds to the impulse approximation.}
\label{rapyps}
\end{figure}

The results of the calculations using Eq.~(\ref{phocs}) are
 presented in Fig.~\ref{shad}. The figure depicts the ratio of 
$\sigma_{\gamma A\rightarrow \Upsilon A}$ calculated with 
Eq.~(\ref{phocs}) to that calculated ignoring the effect of
 nuclear shadowing (setting $G_A=AG_N$). We find a
rather strong suppression of the coherent $\Upsilon$ 
photoproduction cross section off nuclei. 
Note that UPC collisions at LHC are sensitive to  the $x$-range of
$3 \times 10^{-4} \le x \le 5\times 10^{-3}$.

Since the significant effect of shadowing is far beyond the energies available
in the fixed target experiments with the photon beams, the only opportunity
to study this phenomenon is coherent production in UPC at LHC\footnote{In
 principle, HERA can reach into the necessary kinematics.
However, this would require both the approval of HERAIII program and 
development of the system  necessary for high luminosity runs with
heavy nuclei.}. In this kinematics, the coherent events can be selected by the
requirement of anticoincidence with the signal in the  Zero Degree Calorimeter( 
the requirement is that there are   no neutrons in the final state)
and by selecting   the produced $\Upsilon$ with the small transverse momentum,
which strongly suppresses the   contribution of the incoherent diffraction.
The specific transverse momentum distributions for both coherent and incoherent
production are shown in Fig.~\ref{dtyps}. The  cross section of the 
incoherent $\Upsilon$ photoproduction  was estimated in the impulse 
approximation as
\begin{equation}
 {d\sigma_{\gamma A\rightarrow \Upsilon X}(s,t)\over dt}
={A}  
{d \sigma_{\gamma N\to \Upsilon N}(s,t)\over dt} \, .
\end{equation}
This constitutes the upper limit for incoherent contribution.

 The rapidity distributions for coherent $\Upsilon$ production in the
UPC with Ca and Pb beams are shown in Fig.~\ref{rapyps} and the corresponding 
total cross sections are given in Table~\ref{tcrsec}.
For comparison in  Fig.~\ref{rapyps} we also present results (the dotted curve) of
the calculation using the model of Eskola {\it et al.}~\cite{Eskola}
 for the gluon shadowing, which predicts somewhat smaller suppresion of the 
$\Upsilon$ yield. In order to illustrate that that eikonal dipole models of 
rescatterings give a much smaller suppression, we present also the result of 
the eikonal calculation (the dot-dashed curve) using the analysis of the value 
of $\sigma(d,x)$ from~\cite{martin}(the result of \cite{GBW}
 for $d\sim 0.1$ fm are very similar, see comparison in~\cite{martin}).
\begin{table}[]
\begin{center}
\begin{tabular}{|c|c|c|}\hline
 Approximation  & CaCa at LHC($\gamma =3500$)  &PbPb  at 
LHC($\gamma =2760$) \\ \hline\hline
Impulse  &     1.8 $\mu $b         &   133 $\mu $b          \\ \hline
Glauber+Leading twist shadowing   &  1.2 $\mu $b  &   78 $\mu $b       
   \\ \hline
\end{tabular}
\end{center}
\caption{Total cross sections of $\Upsilon$ production
 in UPC at LHC.}
\label{tcrsec}
\end{table}
It appears that a detector with  good acceptance for $\Upsilon$
production for  the currently discussed luminosities 
for heavy ion runs of LHC will collect enough statistics 
to measure the cross section of the discussed process with  good
precision. As  seen from comparison to the calculations in the Impulse Approximation 
(Fig.~\ref{rapyps}), the yield of $\Upsilon$ is expected to be suppressed 
by a factor of two at central rapidities due to  leading twist shadowing.
Practically, the only opportunity to draw the conclusion about the onset 
of the color opacity from the measurement of the $\Upsilon$ yield 
in the ultraperipheral ion collisions at LHC, which we see at the moment,
is the comparison of the calculations with the data. Obviously until
the rather large uncertainty in the absolute value of the 
$\gamma +N\to \Upsilon +N$ cross section  exists, such a conclusion cannot be 
considered as decisive. However, in the coming couple of years it is 
expected that this elementary cross section will be measured at HERA with much
better accuracy. Besides, the energy dependence of the elementary cross 
section can be considered as well established theoretically. Hence, a 
measurement of the
$\Upsilon$ yield at the edge of the rapidity distribution, where the effect
of shadowing is still insignificant, could help to fix the uncertainty and 
verify the calculation of the cross section in the impulse approximation in
all ranges of rapidities. 
Therefore, a comparison with the data at the central
rapidities would provides us  with an estimate of the shadowing effect.    
Note  that there exists a procedure~\cite{Baltz:2002pp} to separate the
production of mesons at small and large impact parameters through the
study of the nucleus break up, which would allow to determine the
ratio of the cross section of $\Upsilon$ production by left and right
moving photons for a given rapidity.
Another possibility will be to measure the elementary cross section at
LHC in UPC collisions of protons with nuclei.
Though this cross section is a factor of $\propto A^{4/3}$ smaller, the
expected luminosity in $pA$ collisions is a factor of 100 larger than in
$AA$  collisions.

\section{Conclusion}

We found that the UPC program at LHC will provide a practical way to
search for the onset of the perturbative color opacity via the study of
the coherent  $\Upsilon$ production  at LHC and will decisively
discriminate between the leading twist and higher twist scenarios
of gluon shadowing at large $Q^2$.

This work was supported by GIF, Sofia Kovalevskaya 
Program of the Alexander von Humboldt  Foundation and DOE.

\end{document}